X-ray photoelectron spectroscopy studies of non-stoichiometric superconducting NbB$_{2+x}$

R. Escamilla and L. Huerta

Instituto de Investigaciones en Materiales, Universidad Nacional Autónoma de México.

04510 México D.F., México

**ABSTRACT**

Polycrystalline samples of NbB$_{2+x}$ with nominal composition (B/Nb) = 2.0, 2.1, 2.2, 2.3, 2.4 and 2.5 were studied by X-ray photoelectron spectroscopy (XPS). The spectra revealed Nb and B oxides on the surface of the samples, mainly B$_2$O$_3$ and Nb$_2$O$_5$. After Ar ion etching the intensity of Nb and B oxides decreased. The Nb 3d$_{5/2}$ and B 1s core levels associated with the chemical states (B/Nb) were identified and they do not change with etching time. The Binding Energy of the Nb 3d$_{5/2}$ and B 1s core levels increase as boron content increases, suggesting a positive chemical shift in the core levels. On the other hand, analysis of Valence Band spectra showed that the contribution of the Nb $4d$ states slightly decreased while the contribution of the B $2p_\pi$ states increased as the boron content increased. As a consequence, the electronic and superconducting properties were substantially modified, in good agreement with band-structure calculations.

**1. INTRODUCTION**

Since the discovery of superconductivity in MgB$_2$ with a transition temperature T$_c$ of 39 K [1] much experimental [2-5] and theoretical [6-8] research has been carried out on this compound and on a series of isostructural diborides. Band structure calculations in MgB$_2$ clearly revealed that while strong covalent B-B bond is retained within boron planes, the Mg-B bond is ionic and the two electrons of Mg are fully donated to the B [9]. On the other hand, studies on the bond ionicity in the $4d$ transition metals diborides have shown that the factor of ionicity (*fi*) of Me-B bond (Me = transition metals) decreased at higher metal atomic numbers (*Z*) [10]. Apparently the existence of delocalization of the valence electrons between layers and other types of bonds induces changes in the stoichiometry and modify the electronic properties in these compounds [11]. Most non-stoichiometric $4d$ transition metal diborides are produced at small *fi's*, such as niobium diboride [12-14].

In spite of the fact that the electronic properties of transition metal diborides have been well studied, details of the electronic structure of non-stoichiometric NbB$_{2+x}$ compounds are a matter of debate in the literature [15, 16]. Moreover, there is no consensus about the character of the chemical bond involved. Some researchers believed that the boron atoms behave as donors [17-19] while others argued that charge transfer occurs in the opposite direction [20-22]. X-ray photoelectron spectroscopy (XPS) is one of the most effective and direct methods to investigate the kind of chemical bonds in molecules and crystalline solids. This paper shows the relevance of the stoichiometry in the electronic and superconducting properties of $NbB_{2+x}$. The chemical state of boron and niobium atoms was estimated.

## 2. EXPERIMENTAL

Samples of $NbB_{2+x}$ were synthesized with nominal composition (B/Nb) = 2.0, 2.1, 2.2, 2.3, 2.4 and 2.5 by the solid-state reaction method. The precursors, commercially available $NbB_2$ powder (Aldrich, -325 mesh) and boron (99.5 % powder, crystalline, < 57 mesh, 99.5 mass %) were mixed in stoichiometric amounts and pressed into pellets of 6 mm in diameter and 0.6–1 g in weight. The pellets were placed in stainless steel sealed tubes and sintered in a tube furnace at 1000 $^o$C for 3 h in an $Ar^+$ atmosphere and quenched to room temperature. Phase identification of the samples was done using an X-ray diffractometer (XRD) Siemens D5000 using Cu-$K_\alpha$ radiation and a Ni filter. Intensities were measured in steps of 0.02$^o$ for 14 seconds in the 2θ range 10$^o$ – 110$^o$ at room temperature. Crystallographic parameters were refined using the program Quanto (A Rietveld program for quantitative phase analysis of polycrystalline mixtures) with multi-phase capability [23]. The chemical analysis was carried out by X-Ray Photoelectron Spectroscopy (XPS). This analysis was performed using an UHV system of VG Microtech ESCA2000 Multilab, with an Mg $K_\alpha$ X-ray source (hν= 1253.6 eV), operated a 15kV and 20 mA beam, and CLAM4 MCD analyser. The surface of the pellets was etched during 5 minutes with 4.5kV $Ar^+$ at 0.33 μA/mm$^2$. The XPS spectrum was obtained at 55° of the normal surface in the constant pass energy mode, $E_0$ = 50 eV and 20 eV for survey and high resolution narrow scan, respectively. The atomic relative sensitivity factor (RSF) reported by Scofield was corrected by transmission function of the analyzer [24] and by reference material $Nb_2O_5$, B, Nb and $B_2O_3$. The peak positions were referenced to the background silver 3$d_{5/2}$ photopeak at 368 eV, having a FWHM of 1 eV, and C 1s hydrocarbon groups in 284.5 eV central peak position. The XPS spectra were fitted with the program SDP v 4.1 [25].

The composition error estimated by XPS is based on the detection limit of the system (0.1%) and the uncertain propagation. In the process of deconvolution the uncertain in the Binding Energy was estimated in 5%. Thus, the uncertain associated to the atomic composition was 3% due to the maximum deviation of the reference materials.

## 3. RESULTS AND DISCUSSION

Fig. 1 shows the powder X-ray diffraction (XRD) patterns obtained for all samples. The main features corresponded to the $NbB_2$ phase (ICDD nº 75-1048). In a previous paper it was shown that [26] a) The most abundant phase was $NbB_2$ having a percentage larger than 94% for all samples, b) Using the Rietveld method the sample composition was estimated $(B/Nb)_{Rietveld}$ and the relation with the nominal composition is indicated in Table 1, c) The boron in excess into the structure is accompanied by the creation of vacancies on the metal (Nb) site, producing important changes in the electronic and superconducting properties and d) The increase of boron induces superconductivity. Recently, C. A. Nunez et al [27] obtained similar results by studies of neutron diffraction in $NbB_{2+x}$ samples.

In order to examine the stoichiometry as well as the formation of some other phases, we analyzed the polycrystalline samples by X-ray photoelectron spectroscopy (XPS). Fig. 2 shows the XPS spectra before (a) and after (b) $Ar^+$ etching for the polycrystalline samples. It is observed that the surface of the polycrystalline samples before etching, exhibits significant levels of C, N and O in addition to the Nb and B oxides. After etching, the intensity of C 1s (Binding Energy, BE = 284.50 eV), N 1s (400 eV) and O 1s (532.00 eV) core levels diminishes.

Figs. 3 a) and 3 b) show the deconvolutioned XPS spectra of the Nb $3d_{5/2}$ band before and after etching for the sample of composition (B/Nb) = 2.0. In the process of deconvoluting of Nb 3d XPS spectra we fixed the values of Binding Energy for $Nb^{5+}3d_{5/2}$, $Nb^{4+}3d_{5/2}$ and $Nb^{2+}3d_{5/2}$ core levels at: 207.57 eV, 206.10 eV and 204.70 eV, respectively.

In fig.3a it might be seen that the sample surface composition is mainly $Nb_2O_5$, which differs qualitatively from the surface after the etching (fig 3b) that presented mainly the $NbB_2$ phase. The presence of $Nb_2O_5$ in the surface might be due to the exposure to the ambient atmosphere as suggested also by other authors [28, 29].

In addition to the core level associated to $Nb_2O_5$, two pairs of core levels of poor intensity were identified. The first one was associated to $NbO_2$ ($Nb^{4+}3d_{5/2}$) and appears at BE = 206.1 eV, 1.47 eV lower than that of $Nb_2O_5$. The second was associated to NbO ($Nb^{2+}3d_{5/2}$) and appeared at BE = 204.7 eV, 2.3 eV above that of the Nb metal (202.40 eV), which is 0.4 eV on average below the value previously reported [30]. After $Ar^+$ etching, the intensity of $NbO_2$ and $NbO$ peaks increased, in particular, we observed that the intensity of $NbO$ core levels is higher than that associated to $NbO_2$ [31].

Fig.4 shows the effect of the etching time on the core levels energy for the sample of composition (B/Nb) = 2.0. It might be seen that there is a significant chemical shift after etching time as short as 3 min. This indicated the erosion of the $Nb_2O_5$ phase leaving a nearly pure $NbB_2$ phase after 5 min. Comparing the BE of the Nb $3d_{5/2}$ peak after 5min of etching with the Nb metallic reference we observed a 0.94 eV chemical shift. As we can see, for etching time longer that 5 min there was no further changes in the Nb 3d core level position, suggesting that the stoichmetry of the samples remained stable.

In order to determine the sample compositions, the atomic concentration was calculated by XPS using the survey spectra and RSF of Nb 3d (8.210) and B 1s (0.486) obtained by the reference samples. As can be observed in Table 1, the calculated compositions by XPS are very close to the compositions calculated by Rietveld refinement method [26] (see Table 1). Table 1 show, apart from the composition of the samples estimated by three different methods, the B 1s and Nb $3d_{5/2}$ energy position and the critical superconducting temperature.

Fig. 5 a) shows the deconvolution of the XPS spectra in the Nb 3d region for the different samples after 5 min of etching. From the fitting, we observed an increase of BE of the Nb $3d_{5/2}$ for samples in the composition range $2.00(6) \leq (B/Nb)_{XPS} \leq 2.27(7)$ and a slight decrease for the $2.31(7) \leq (B/Nb)_{XPS} \leq 2.44(8)$. Therefore, we observed a positive chemical shift in the Nb $3d_{5/2}$ core level with respect to the Nb metallic for all samples. Similar positive chemical shift are observed in the 3d transition metal borides respect to metals [32].

Fig. 5 b) shows the deconvolution of the XPS spectra in the B 1s region for all samples. The B 1s core level associated with the sample of composition $(B/Nb)_{XPS} = 2.00(6)$ was localized at BE = 188.15 eV, this value was within the binding energy variation range for typical transition metal diborides [33, 34] and borocarbides $RNi_2B_2C$ (R =Y and La) [35]. (187.1 – 188.3 eV). This observation is consistent with reported calculations and maximum entropy method (MEM) results, which have shown that B – B bonding is two-dimensionally covalent ($sp^2$) [6, 36-38]. Furthermore, it was observed that there is an increase of the BE of B 1s core level in the range $2.00(6) \leq (B/Nb)_{XPS} \leq 2.27(7)$ and it remained constant for greater concentrations (see Table 1). In all samples a positive chemical shift was calculated with respect to the boron reference sample, contrary to what have been observed for 3d transition metal borides [32]. The maximum positive chemical shift (0.90 eV) was obtained for the compositions $(B/Nb)_{XPS} = 2.27(7)$, 2.31(7) and 2.44(7). In addition, core levels at BE = 193.10 eV, 195.10 eV and 197.67 eV were observed, the first one correspond to $B_2O_3$ while the rest are associated to satellites of Nb due to the x ray source (Mg $K_\alpha$) and not to satellite shake-up of boron compounds [39].

It is important to point out that the chemical shifts in BE are often used to study the electronic redistribution or charge transfer upon compounds and alloys. In a conventional XPS interpretation, the general rule is that the BE of the central atom increases as the electronegativity of the attached atoms or groups increases [40]. Since B (2.04) is more electronegative than Nb (1.6) according to Pauling's electronegativity table [41], one would expect that the B core level shifts toward lower binding energy. As has been observed for $TiB_2$; the BE of B1s is lower that in pure boron and the BE of Ti $2p_{3/2}$ is higher that Ti metal, in this case the authors assumed that some charge transfer occurred from the Ti atoms to the boron atoms.

This observation was confirmed by results obtained by the discrete-variational $X\alpha$ method [42]. The magnitude of the electron donation decreased from $ScB_2$ to $FeB_2$; the higher donor ability of scandium and the smaller donor ability of titanium in the diborides were corroborated by the fact that the BE of the B 1s in $TiB_2$ was greater than that in $ScB_2$ and even smaller than that in pure boron [43].

Furthermore, the studies of XPS in $MgB_2$ have shown that the BE of B1s is lower than that in pure boron [44, 45] and the BE of Mg 2p is higher than that in Mg metal [45], suggesting that some charge transfer occurs from the Mg atoms to the boron atoms. Studies of the Valence-electron distribution in $MgB_2$ by accurate diffraction measurements and first-principles calculations confirmed this observation [46]. From the data in Table 1, can be observed that the general rule based on the electronegativity fails to explain the positive chemical shift of the B 1s core level measured in this work. Therefore, the presence of superconductivity in these samples cannot be explained by a charge transfer model based only on chemical shift effects.

In order to determine the effect of boron excess on the density of states at the Fermi level $N(E_F)$, we measured the Valence Band spectra using a monochromatic Al K$\alpha$ source. Fig. 6 shows the normalized Valence Band spectra for the sample of compositions (a) $(B/Nb)_{XPS} = 2.00(6)$ and (b) $(B/Nb)_{XPS} = 2.44(8)$ compared with the (c) total density of states (DOS) determined from band-structure calculations [47]. The discontinuous lines delimit the niobium and boron states respect to $N(E_F)$. A good correspondence between experiment and theory can be obtained if DOS is shifted to lower binding energy, as has been done for high –temperature superconductors, where a $\sim 2$ eV shift was required, and was attributed to electron correlation effects.

If the shifts done, the feature between 8 and 12 eV is due to the B $2s$ states observed also in MgB$_2$ [7,9, 49,50] while the feature between 5–8 eV is predominantly due to B $2p_\pi$ states [48]. Whereas, the feature around 2 eV is associated to the Nb $4d$ states [15]. Because the main contribution to the total density of states at the Fermi level of NbB$_2$ are the Nb $4d$ states, the $N(E_F)$ of this phase is greater than that for MgB$_2$, the former is 1.074 states/eV-cell while that for the latter is 0.719 states/eV–cell [47].

Comparing the Valence Band spectra for both compositions, we observed that for $(B/Nb)_{XPS} = 2.44(8)$, the contribution of the B $2p_\pi$ states increases and a slight decrease in the contribution of the Nb $4d$ states is observed, respect to $(B/Nb)_{XPS} = 2.00(6)$.

On the other hand, in a previous paper it was shown [26] that the samples of composition $(B/Nb)_{XPS} \geq 2.15(6)$ display superconductivity reach the maximum $T_c$ at about 9.75 K for a composition $(B/Nb)_{XPS} = 2.44(7)$ (see Fig.7).

Therefore, we supposed that the increase in the $T_c$ in our samples can be explained by the increases in the carrier density (electrons) to the in-plane conduction due to an increase in the number of niobium vacancies that are produced by the boron excess [14]. Muon spin rotation/relaxation measurements in NbB$_{2+x}$ samples confirm the increase of the carrier density as boron content increases [51]. As result of increases of carrier density an increase in the Valence Band due to B $2p_\pi$ is observed.

## 4. CONCLUSIONS

We have prepared samples of NbB$_{2+x}$ with nominal composition ranging from $(B/Nb) = 2.0$ to 2.5 by the solid-state reaction method. The sample compositions were calculated using XPS and a good correlation with the compositions calculated by Rietveld refinement method was obtained. The stoichiometry of these compounds was stable during large periods of Ar$^+$ etching time. Particularly, variations in the core level energies of the Nb 3d and B 1s peaks were correlated to the different sample compositions. For the sample of composition $(B/Nb)_{XPS} = 2.00(6)$, the Nb $3d_{5/2}$ and B 1s core levels are localized at 203.34 eV and 188.15 eV, respectively. As a consequence of the increase in the boron content, a positive chemical shift was observed in the Nb $3d_{5/2}$ and the B1s core levels. For the compositions $(B/Nb)_{XPS} \geq 2.27(7)$ we observed the maximum positive chemical shifts in the Nb $3d_{5/2}$ and B 1s core levels. The study of the Valence Band for the $NbB_2$ phase was consistent with band-structure calculations, our results showed a slight decrease in the contribution of the Nb 4$d$ states and an increase in the contribution of the B 2$p_\pi$ states to the density of states at the Fermi level $N(E_F)$ with the increase of boron content. Finally, we observed that the charge transfer model based on the concept of electronegativity was not applicable to explain the superconductivity in the $NbB_{2+x}$ samples. However, we could associated the maximum $T_c$ (9.4 – 9.75 K) to the increment in the carrier density due to major contribution of B 2$p_\pi$ electrons to the Valence Band.

## ACKNOWLEDGEMENTS

Support from DGAPA-UNAM under project PAPIIT-IX101104 and PAPIIT IN119806-2 is gratefully acknowledged. Thanks to S. Rodil and J. Alonso for carefully reading and correcting the manuscript.

**FIGURE CAPTIONS**

Fig. 1. X - ray diffraction patterns for all samples.

Fig. 2. XPS spectra for all samples before (a) and (b) after etching. The arrows indicate Nb, N, O, C and B states.

Fig. 3. XPS spectra Nb 3d of (B/Nb) = 2.0 before (a) and (b) after Ar[+] etching. The arrows indicate Nb states.

Fig. 4. XPS spectra Nb 3d and B 1s as a function of the Ar[+] etching time for (B/Nb) = 2.0.

Fig. 5. XPS spectra for (a) Nb 3d and (b) B 1s for $(B/Nb)_{XPS}$ = 2.00(6), 2.14(6), 2.15(6), 2.27(7), 2.31(7) and 2.44(8) after 5 min Ar[+] etching. The points represent the experimental spectrum and the line represents the result of the deconvolution.

Fig. 6. Comparison of the measured $(B/Nb)_{XPS}$ = 2.00(6) and 2.44(8) Valence-Band spectra with the total density of states (DOS) of $NbB_2$ calculated [47].

Fig. 7. Superconducting transition $T_c$ as a function of the composition $(B/Nb)_{XPS}$

**TABLE CAPTIONS**

Table 1. (B/Nb) Nominal composition, $(B/Nb)_{Rietveld}$ composition obtained by Rietveld refinement method, $(B/Nb)_{XPS}$ composition obtained by XPS, B 1s and Nb $3d_{5/2}$ positions of the spectral lines, $\Delta$B 1s and $\Delta$Nb $3d_{5/2}$ shift chemical, Superconducting transition $T_c$.

Figure1

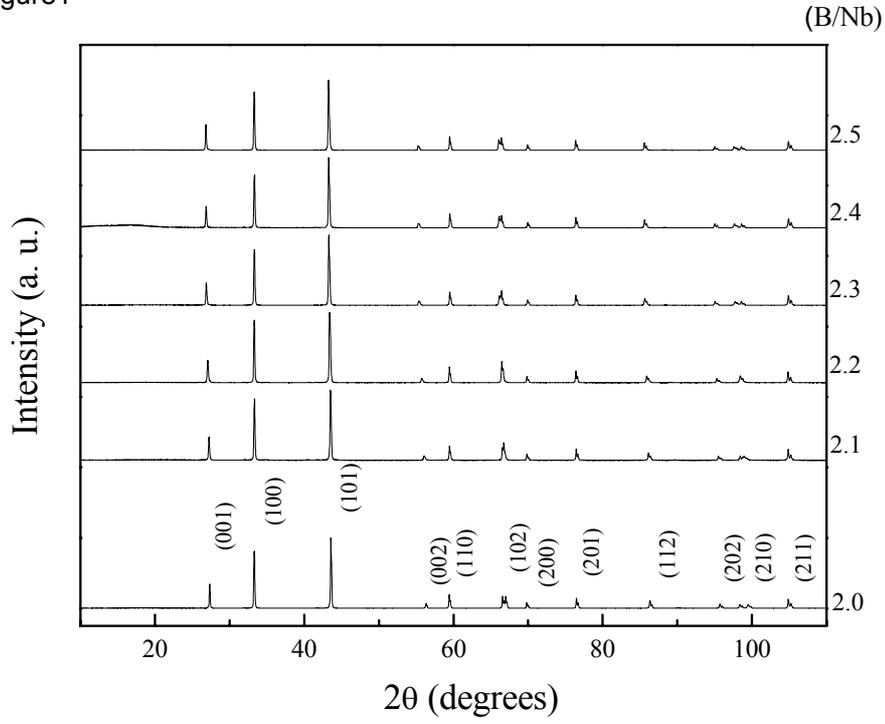

Figure2

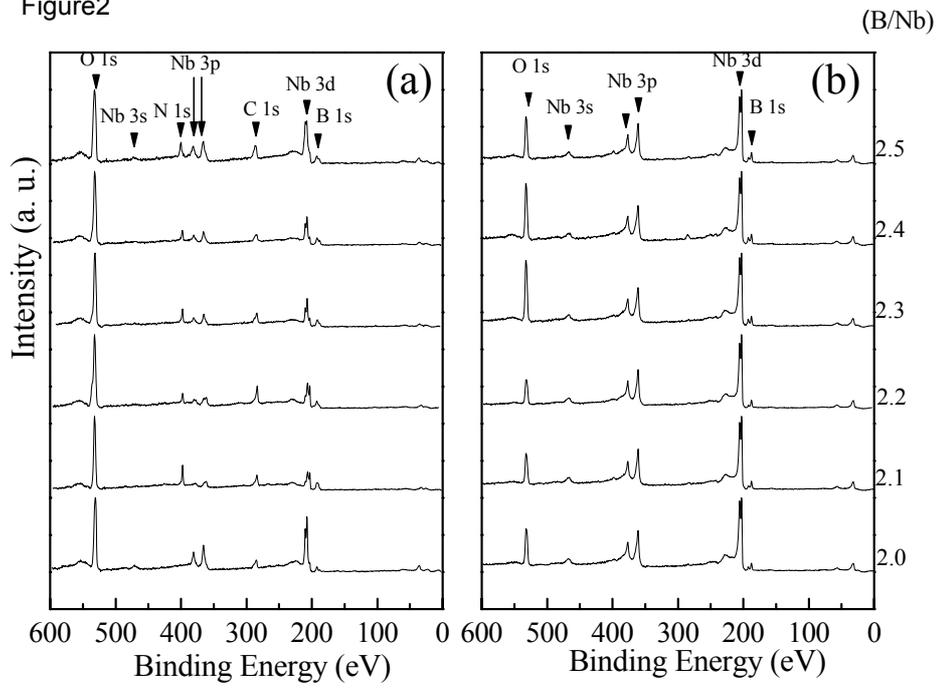

Figure3

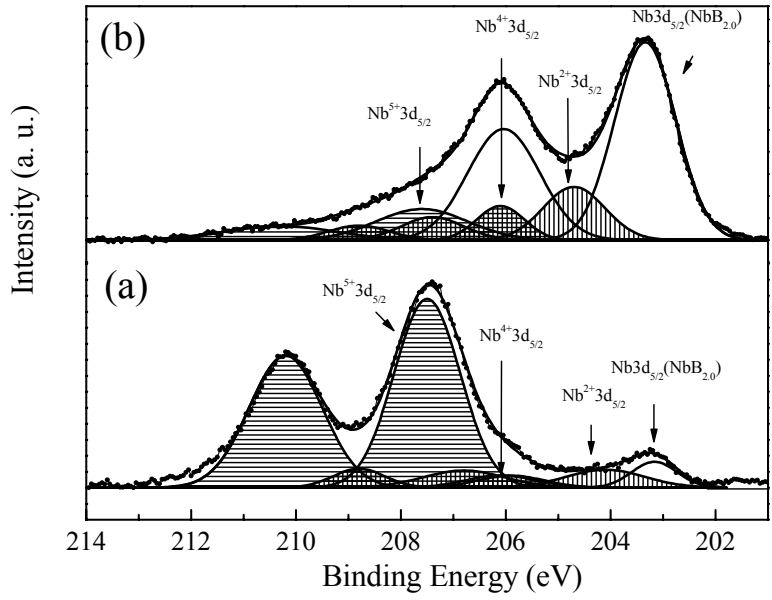

(b)

Nb⁵⁺3d₅/₂  →  Nb⁴⁺3d₅/₂  →  Nb²⁺3d₅/₂  →  Nb3d₅/₂(NbB₂.₀)

(a)

Nb⁵⁺3d₅/₂  →  Nb⁴⁺3d₅/₂  →  Nb³⁺3d₅/₂  →  Nb3d₅/₂(NbB₂.₀)

Binding Energy (eV)

Figure4

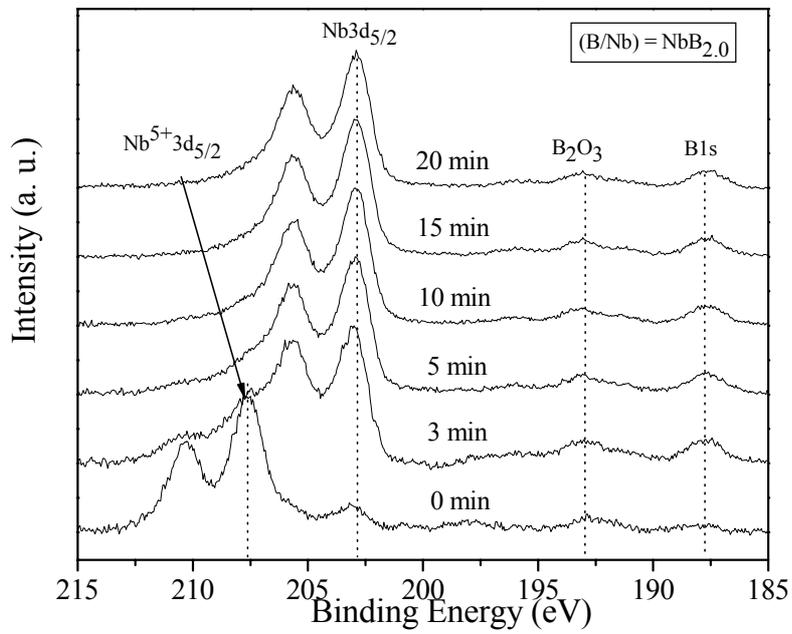

(B/Nb) = NbB₂.₀

Nb3d₅/₂

Nb⁵⁺3d₅/₂

B₂O₃   B1s

20 min
15 min
10 min
5 min
3 min
0 min

Binding Energy (eV)

Figure5

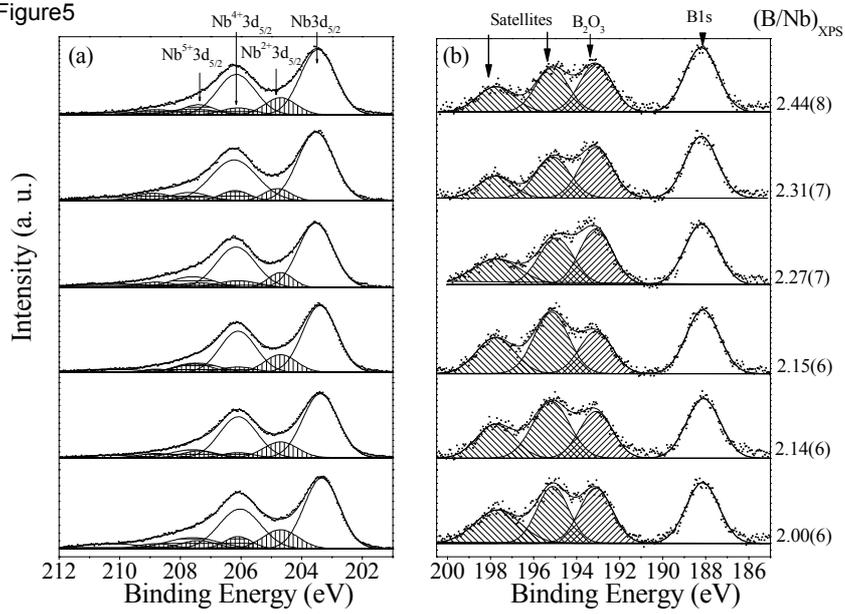

Figure6

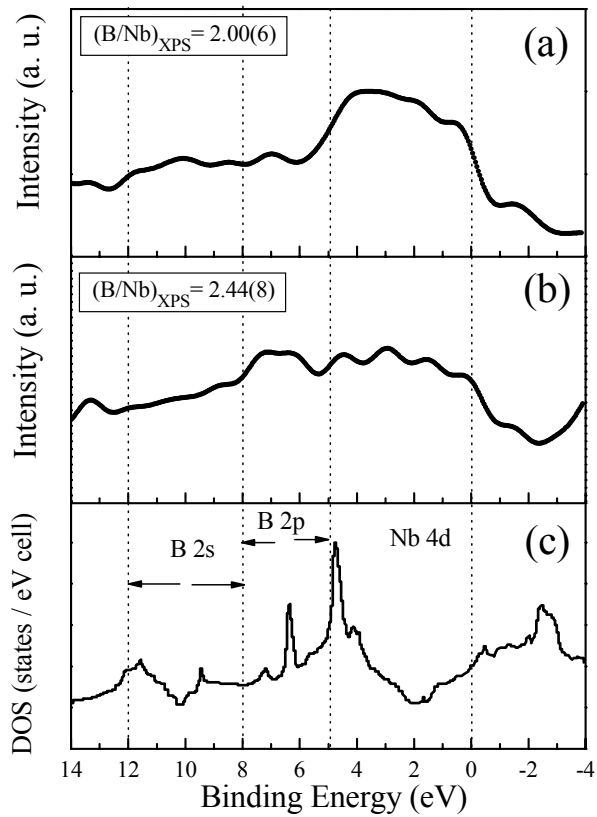

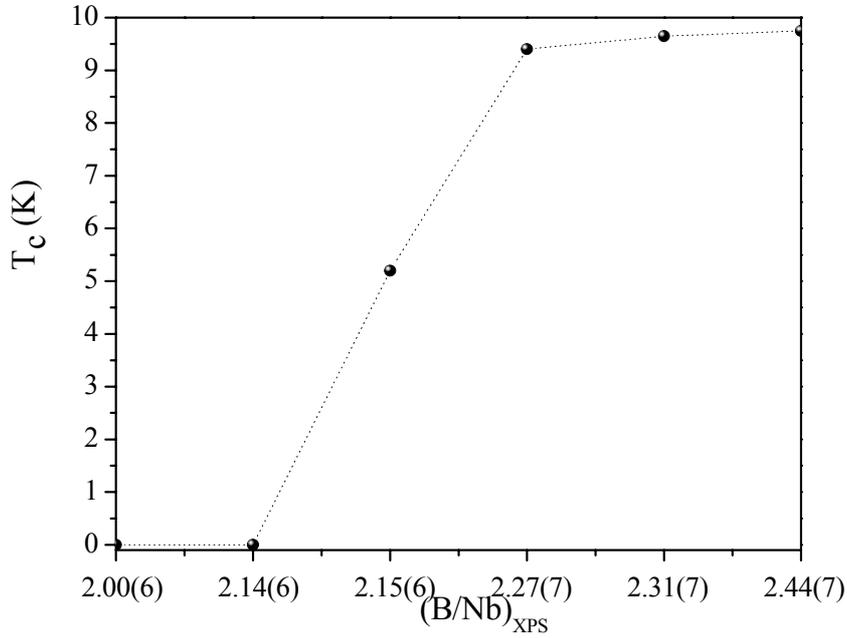

Figure7

**TABLE 1**

| (B/Nb) | (B/Nb)$_{Rietveld}$ | (B/Nb)$_{XPS}$ ±3% | B 1s (eV) | ΔB 1s (eV) | Nb 3d$_{5/2}$ (eV) | ΔNb 3d$_{5/2}$ (eV) | T$_c$(K) ±0.05 |
|---|---|---|---|---|---|---|---|
| 2.0 | 2.00(1) | 2.00 | 188.15 | 0.85 | 203.34 | 0.94 | 0 |
| 2.1 | 2.10(1) | 2.14 | 188.11 | 0.81 | 203.39 | 0.99 | 0 |
| 2.2 | 2.20(2) | 2.15 | 188.13 | 0.83 | 203.40 | 1.00 | 5.20 |
| 2.3 | 2.30(1) | 2.27 | 188.20 | 0.90 | 203.54 | 1.14 | 9.40 |
| 2.4 | 2.32(1) | 2.31 | 188.20 | 0.90 | 203.53 | 1.13 | 9.65 |
| 2.5 | 2.34(1) | 2.44 | 188.20 | 0.90 | 203.46 | 1.06 | 9.75 |